\documentclass[prl, superscriptaddress, twocolumn, showpacs]{revtex4}

\usepackage{amsfonts}
\usepackage{amssymb}
\usepackage[dvips]{graphicx}
\usepackage{color}
\usepackage{amsmath}
\usepackage[bookmarksnumbered, bookmarks, breaklinks, linktocpage]{hyperref}

\newcommand{\eend}      {\hspace{\stretch{1}}\rule{1ex}{1ex}}

\begin{document}

\title{Dynamics of a Quantum Phase Transition}

\author{Wojciech H. Zurek}
\affiliation{Theory Division, LANL, MS-B213, Los Alamos, NM  87545, USA}
\author{Uwe Dorner}
\affiliation{Clarendon Laboratory, University of Oxford, Parks Road,
  Oxford OX1 3PU, United Kingdom}
\author{Peter Zoller}
\affiliation{Institute for Theoretical Physics, University of Innsbruck,
and Institute for Quantum Optics and Quantum Information of the
Austrian Academy of Sciences, A-6020 Innsbruck, Austria}

\date{\today}

\begin{abstract}
We present two approaches to the dynamics of a quench-induced
phase transition in the quantum Ising model. One follows
the standard treatement 
of 
thermodynamic second order phase transitions
but applies it to the quantum phase transitions. The other approach is
quantum, and uses Landau-Zener formula for transition
probabilities in avoided level crossings. We show that predictions of the two
approaches of how the density of defects scales with the quench rate
are compatible, and discuss the ensuing insights into the dynamics of
quantum phase transitions.
\end{abstract}

\maketitle

Studies of phase transitions traditionally focussed on {\it
equilibrium} scalings of various properties near the
critical point. The first major exception was an attempt to model
the physics of the early Universe:
Kibble~\cite{Kib76} noted that cosmological phase transitions in a
variety of field theoretic models lead to formation of topological
defects (such as monopoles or cosmic strings) which may have
observable consequences. One of us then pointed out \cite{Zur85a}
that analogues of cosmological phase transitions can be studied in
the laboratory. In such experiments the {\it equilibrium} critical scalings
predict various aspects of the {\it non-equilibrium} dynamics of
symmetry breaking, including the density of residual topological defects
\cite{Zur85a, Zur96a}.

These ideas 
led to the {\it Kibble-Zurek mechanism} (KZM), a theory
of defect formation that uses the critical scalings of the relaxation time 
and of the healing length to deduce size ($\hat \xi$) of domains that
choose the same ``broken symmetry vacuum'' \cite{Zur96a, Kib03}.
When the broken symmetry phase permits their existence, KZM predicts 
defects will appear with density of about one defect unit (e.g., one
monopole or a $\hat \xi$-sized section of a string) per 
$\hat \xi$-sized domain. This KZM prediction has been tested, extended and
refined with the help of numerical simulations \cite{LZ96a,
RH00a}, and verified in a variety of increasingly sophisticated
and reliable experiments in liquid crystals \cite{Chu92a, Bow96a},
superfluids \cite{He4a, He4b, He3}, superconductors \cite{Mon00a,
Car00, Man03}, and other systems \cite{Are00}.

A majority of the experimental data agree with KZM. One notable
exception is the case of  superfluid 4He, where initial reports of
KZM vortices being detected \cite{He4a} were retracted \cite{He4b}
after it turned out that stirring had inadvertently induced 
vorticity. 
In view of various uncertainties, it is still not clear whether 4He 
experiments are at odds with the numerics-assisted KZM predictions.
Regardless, KZM provides a theory of the dynamics of second 
order phase transitions ranging from low temperature Bose-Einstein 
condensation to grand unification scales encountered in particle physics and cosmology.

In this paper we consider a barely explored problem: the
dynamics of quantum phase transitions. Quantum many-body systems
(e.g., Bose gases) can undergo thermodynamic phase transformation (such as
Bose-Einstein condensation that follows evaporative cooling). KZM theory,
developed to deal with thermodynamic phase transitions, applies in this case directly,
even though the dynamics of Bose condensation is explicitly quantum \cite{AZ99a}.

On the other hand, a quantum phase transition \cite{Sac99a}
(e.g., the Mott insulator-superfluid transition of bosons in a periodic lattice) 
is a change in the character of a system's {\it ground state} which occurs as
some parameter of its Hamiltonian passes critical value. For instance, lowering 
of the amplitude of the optical lattice induces Mott transition.
Unlike thermodynamic transitions, quantum phase transitions 
involve only reversible unitary dynamics. Therefore, scaling arguments
that work in thermodynamic transitions (where the order
parameter is damped) may not be valid in the quantum case
(but see \cite{DSBZ02,Dor03}). Yet, dynamics of quantum phase
transitions is interesting in its own right and has
applications in quantum information processing \cite{Dor03, Cal04}.

We will study a quench-induced transition in the quantum Ising model. 
This model is regarded by Sachdev \cite{Sac99a} as one of two canonical 
quantum phase transitions. It describes  a chain 
of spins with the Hamiltonian:
\begin{equation}
H=-J(t) \sum_{l=1}^N \sigma_l^x - W\sum_{l=1}^{N-1} \sigma_l^z \sigma_{l+1}^z \ .
\label{eq:Hamiltonian}
\end{equation}
Here $\sigma_l^{x},~\sigma_l^{z}$ are Pauli operators, $W$ is the Ising coupling,
and $J(t)$ is proportional to  the strength of an external field that attempts
to align  spins with the $x$-axis.

The phase transition from the high-field state (all spins aligned with 
$x$, i.e., $|\rightarrow,\rightarrow, \dots, \rightarrow \rangle$) to
the low-field ground state manifold -- spanned by 
$|\uparrow, \uparrow,\dots, \uparrow\rangle$ 
and $ |\downarrow, \downarrow, \dots \downarrow\rangle$, and doubly degenerate 
in the large $N$ limit -- takes place when $J(t)=W$. Thus, {\it relative coupling}:
\begin{equation}
\epsilon(t)={J(t) / W} - 1
\label{eq:epsilon}
\end{equation}
is expected to play a role of the relative temperature
$({T(t) \over T_C} - 1)$ in system's behavior near the critical
point $T_C$. 

Indeed, all relevant properties depend on the
size of the {\it gap} ($\Delta$) between the ground state and the
first excited state. As $N \rightarrow \infty$, the gap is:
\begin{equation}
\Delta=2|J(t)-W|=2W|\epsilon(t)| \ .
\label{eq:gap}
\end{equation}
and sets an energy scale reflected in the {\it relaxation time}
\begin{equation}
\tau={\hbar /  \Delta} = {\hbar / {2 W| \epsilon(t)| }}={\tau_0 / |\epsilon(t)| } \ .
\label{eq:tau}
\end{equation}
Divergence of $\tau$ near the critical point is the {\it
critical slowing down} familiar from thermodynamic phase transitions. 

The {\it healing length} 
is given by the product of the
speed of sound $(c)$ and the relaxation time:
\begin{equation}
\xi=2Wa/\Delta(t) = a / |\epsilon(t)| = \xi_0/|\epsilon(t)| \ ,
\label{eq:xi}
\end{equation}
where $c=2Wa/\hbar$ (see \cite{Sac99a}),
and $a$ is the distance between spins. The divergence of  
$\xi$ near the critical point is analogous to {\it critical opalescence}.

The scaling of $\tau$ and $\xi$ suggests estimating the size 
of broken symmetry domains (i.e., regions of aligned spins)
using the same approach that worked in thermodynamic transitions
\cite{Zur85a,Zur96a}: Near the critical point 
``reflexes''  of the system (measured by the relaxation time $\tau$) 
deteriorate, until -- at the critical point, where $\tau=\infty$ -- 
system
cannot react at all. Yet, early in the quench
$\tau$ is still small, and
its state is still able to adjust to variations of the external parameter 
(e.g., $J$ or $T$). This suggests splitting the quench into the near-critical 
impulse regime and the quasi-adiabatic regime far from the critical point. 
Such split is the essence of KZM. 

The instant ($\hat t$) when behavior of the system changes from adiabatic to
impulse is of key importance. This happens when its reaction 
time (given by Eq.~(\ref{eq:tau})) is the same as the timescale on which 
its Hamiltonian is changed. To calculate $\hat t$, we assume that 
the external bias field changes linearly with time, so that $\epsilon(t)= t/\tau_Q$. 
As the relative coupling changes on a timescale $\epsilon(t)/\dot\epsilon(t) =t$, 
the switch between adiabatic and impulse regimes occurs at the instants
$\pm \hat t$ when relaxation time is equal to $t$,
\begin{equation}
\tau(\hat t) = \ {\tau_0 / | \epsilon(\hat t)| } \  = \ \tau_0\tau_Q/\hat t 
\  = \  {{ \epsilon(\hat t)} / {\dot \epsilon(\hat t)}} = \hat t
\label{eq:thateq}
\end{equation}
which yields 
\begin{equation}
\hat t = \sqrt {\tau_Q \tau_0} =  \sqrt{{\tau_Q \hbar} / {2 W}} \ .
\label{eq:that}
\end{equation}

Typically, these two instants ($\pm \hat t$) separate evolution into three regimes. 
Initially, for $t<-\hat t$, the system's state will adjust 
to the decreasing $J(t)$. However, at $t= - \hat t$ (before the critical
point) this tracking of the instantaneous ground state of $H$ will cease. Evolution
will re-start only at $+\hat t$ (after the critical point), with an initial state similar to
the one ``frozen out'' at $- \hat t$.

In thermodynamic phase transitions fluctuations of the order
parameter at $\hat t$ give rise to domains of size $\hat \xi$
given by the healing length at $- \hat t$. Using
the relative coupling $\hat \epsilon$ at $\hat t$ we similarly
calculate for the quantum Ising model:
\begin{eqnarray}
&& \hat \epsilon  \equiv   \epsilon(\hat t)  ={ {\hat t} /  \tau_Q} = \sqrt{ \tau_0 /  \tau_Q};
\label{eq:epshat}\\
&& \hat \xi  \equiv \xi_0/{\hat \epsilon} =  \xi_0 \sqrt{ \tau_Q / \tau_0} = a \sqrt{{2W \tau_Q} / {\hbar}} \ .
\label{eq:xihat}
\end{eqnarray}
Note that this scaling differs from the $\hat \xi = \xi_0/\sqrt{\hat
\epsilon} = \xi_0 (\tau_Q/\tau_0)^{1 \over 4}$ predicted by
non-relativistic mean-field theories for second order
phase transitions \cite{Zur85a,Zur96a}.

Following KZM, we now expect appearance of $O(1)$ defects per
$\hat \xi$.  Their density should be therefore:
\begin{equation}
\hat \nu_{KZM} \simeq  a /  \hat \xi =  \sqrt{ {\hbar}  /  {2W \tau_Q}  }.
\label{eq:Nhat}
\end{equation}
This is only an estimate. Simulations of classical second
order transitions yield defect densities that scale with $\tau_Q$
as predicted by KZM, but that can be lower by about an order of magnitude
than $a/\hat \xi$:
Defects can be separated not by $\hat \xi$ but by approximately 10~$\hat \xi$
(see  \cite{LZ96a}).

\begin{figure}[tp]
\centering
\includegraphics{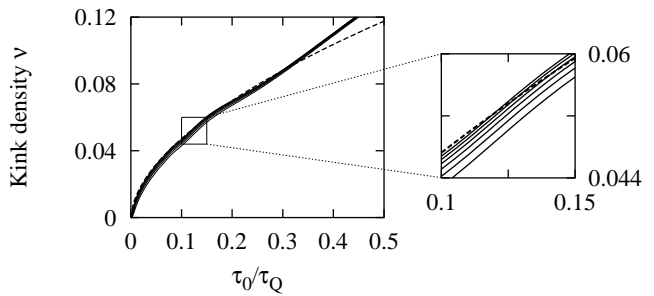}
\caption{Density of kinks ($\nu$) in quantum Ising model after a
quench that starts in a ground state at $J=5W$ and ends at $J=0$,
plotted versus the dimensionless quench rate
$\tau_0/\tau_Q = \hbar v/4W^2$ for $N=50,60,70,80,90,100$ (solid
lines; bottom to top). The simulation data are consistent with the scaling 
$\hat \nu_{KZM} \sim  \sqrt {\tau_0 / \tau_Q}$, Eq. (\ref{eq:Nhat}),
predicted by the
Kibble-Zurek mechanism. (See
\cite{Dor03} for details on the numerical method). Agreement
improves with $N$; for 100 spins, a fit gives
$\nu \sim \tau_Q^{-0.58}$ (dashed line). As in the classical
case~\cite{LZ96a} Eq. (\ref{eq:Nhat}) is an overestimate; the best
fit is $\nu \simeq 0.16 ~ \hat \nu_{KZM}$.} \label{nu}
\end{figure}
\begin{figure*}[tp]
\centering
\includegraphics{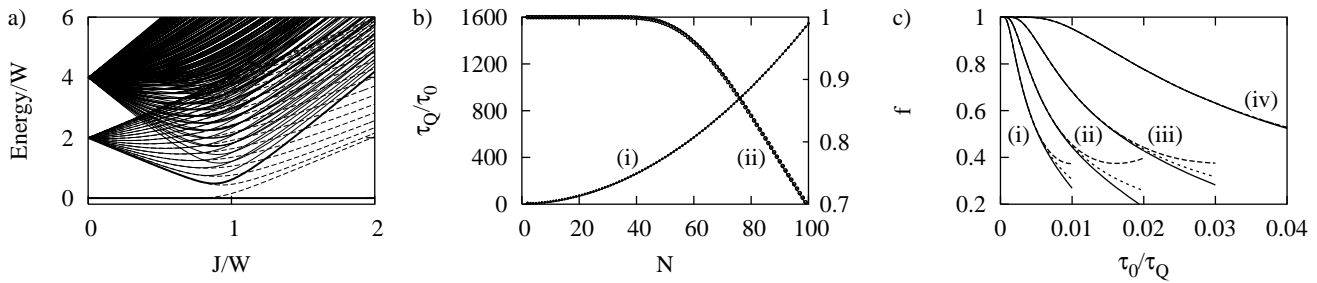}
\caption{(a) Energies of lowest excitations (see~\cite{Dor03})
for $N=20$. The energies of the ground state and the first {\it accessible} excited state are the lowest (horizontal) solid line and the second lowest solid line. (b) (i) Quench time $\tau_Q
/\tau_0=4W^2/\hbar v$ that yields $f$ of 99\%, and (ii) the
fidelity for a fixed $\tau_Q =200\hbar/W=400\tau_0$ as a function
of the number of spins $N$ in the quantum Ising chain. A power-law
fit to the data corresponding to $\tau_{Q_{99\%} }$ gives a power
of 1.93 (LZF yields 2, as would  KZM). The best fit for the
fidelity with Landau-Zener dependence $f=1-\exp{\{{{-a W\tau_Q }
/{\hbar N^2}}\}}$ yields $a\simeq 59$, compared to
theoretical $a=2 \pi^3\simeq 62$ based on the steeper slope in the marked energy level above ($a=4\pi^3\simeq124$ when the more shallow slope in Fig.~\ref{lzf}(a) consistent 
with Eq. (3)  is taken). 
(c) Upper (dashed lines) and lower (solid lines) 
bounds on fidelity as a function of $\tau_0/\tau_Q = \hbar v/4W^2$ for
$N=90$ (i), $N=70$ (ii), $N=50$ (iii), $N=30$ (iv). Fits to LZF (dotted lines) lie between
these bounds, and for $f>0.6$ give $a\simeq 59$ (i), $59$ (ii), $57$ (iii),
$54$ (iv).}
\label{lzf}
\end{figure*}

This KZM paradigm should not be uncritically applied to quantum phase 
transitions. Above all, thermodynamic fluctuations are `real'. If they
survive from 
before 
-$\hat t$, they can tip the balance at $+\hat t$, breaking 
symmetry right after the transition. It is hard to make an analogous 
argument for a quantum case equally convincing. Quantum fluctuations, exist, 
but they are virtual, so it is not obvious that they will have a similar
symmetry breaking 
effect on the post-transition state. On a more prosaic note [as we shall 
see in Fig.~\ref{lzf}(a)] the {\it relevant} gap (i.e., gap between the ground 
state and the first {\it accessible} excited state) is not the symmetric $\Delta$, 
Eq.~(\ref{eq:gap}). Rather, 
its
slope is twice as large on the approaching side.

Nevertheless, Fig.~\ref{nu} shows that number of kinks per spin -- the residual kink density 
created by quenching the quantum Ising model scales approximately as
$ \sim 1/\sqrt{\tau_Q}$, just as predicted by Eq.~(\ref{eq:Nhat}). This holds throughout
the region of KZM's validity, i.e. where $\hat \epsilon$ is much less than 1 (so that 
the quench is quasi - adiabatic early on and at the end, but  `impulse - like'
near the critical point, and thus at least one defect is expected). The
prefactor $\sim 0.16$ [0.12 if the steeper slope on the approach 
in Fig.~\ref{lzf}(a) is taken] is also not far from previous experience
\cite{LZ96a}. 

Kibble-Zurek mechanism works in a quantum transition! Yet, in view of 
doubts about quantum fluctuations, an explicitly quantum treatment would 
be reassuring.

As  $N \rightarrow \infty$, the gap  ($\Delta$) -- a salient feature of the 
quantum Ising model --
disappears at the critical point in accord with Eq. (\ref{eq:gap}). When $N < \infty$, 
this critical gap is small, but finite [see Fig.~2(a)]. 
This is of key importance for the remainder of our paper.
Instead of calculating the density of defects in an infinite system, we shall
compute size ($\tilde N$) of the largest spin chain likely to remain defect free (i.e., 
in a ground state) after a quench, as a function of quench timescale $\tau_Q$. 
For defect probability of $\sim$50\%, inverse of $\tilde N$ is an estimate of defect
density. 

Excited eigenstates of $H$, 
Eq.~(\ref{eq:Hamiltonian}), 
on the broken symmetry ($W>J$) side of the transition describe
states of the spin chain in which the direction of symmetry breaking
varies along the chain once, twice, etc. \cite{Sac99a}. Thus, they represent
states containing one, two, etc. ``kinks''. The behavior of the energies
of lowest excitations of $H$ in the vicinity of the critical point,
[Fig.~\ref{lzf}(a)], suggests avoided level crossing. Hence, it
appears that phase transition dynamics in the quantum Ising model
can be treated using the Landau-Zener formula \cite{LZ}, or LZF. LZF gives
the probability of exciting a system driven through an avoided
level crossing:
\begin{equation}
p_{CHANGE} \simeq \exp\bigl( -{{\pi \hat  \Delta^2} \over {2 \hbar |v|}}\bigr) \ .
\label{eq:LZF}
\end{equation}
Here, $\hat \Delta$ is the minimum energy gap between the two levels, 
and $v$ is the quench velocity. That is, far away from the ``point of 
the nearest approach",  $v=\dot \Delta$.

Using LZF we can compute the average size $\tilde N$ of a spin chain
that is likely to remain in the ground state throughout the
quench. In the adiabatic limit ($v \approx 0$), Eq. (\ref{eq:LZF})
predicts that the system will stay in the same energy eigenstate 
(i.e., the probability of switching levels will be vanishingly small). 
To quantify this Ref. \cite{Dor03} uses the fidelity, $f = |\langle \psi_{ACTUAL} |
\psi_{GROUND}\rangle|^2$, which gives the probability that no defects 
will be produced. From LZF it follows that $p_{CHANGE} \simeq 
\exp\bigl( -{{\pi \hat  \Delta^2} \over {2 \hbar |v|}}\bigr) \simeq 1-f \ $. 
Thus, the rate of a nearly defect - free quench (resulting in 
defects with probability 1-$f  \ll 1$) is bounded:
\begin{equation}
|v| \leq {{\pi \hat \Delta^2} \over {2 \hbar |\ln(1-f)|}} \ .
\label{eq:speed}
\end{equation}
Below we will express $v$ using quench time $\tau_Q$ and $W$; $v=|\dot \Delta|
= 2\dot J(t)={2W/ \tau_Q}$.

The lowest excited states are inaccessible -- they have a different parity than
the ground state, and $H$ conserves parity. The first {\it accessible} level 
has one kink for $J=0$. It gets to within
$\hat \Delta = 4 \pi W / N$ for $N\gg1$ above the ground state.
With these ingredients using Eq.~(\ref{eq:speed}) we obtain:
\begin{equation}
|v| =|\dot \Delta| = {{2 W} \over \tau_Q}  \leq {{\pi (4 \pi W/ \tilde N)^2} \over {2 \hbar |\ln(1-f)|}} \ .
\end{equation}
It relates the size $\tilde N$ of a defect-free chain to quench time:
\begin{equation}
\tilde N \leq  {2 \pi} \sqrt{{\pi W \tau_Q} \over {\hbar |\ln(1-f)|}} = {2 \pi W}
\sqrt{{2 \pi} \over {\hbar v |\ln(1-f)|}} \ .
\label{eq:ntilde}
\end{equation}
Figs.~2(b) and 2(c) show that LZF provides a good fit for $f >0.5$.  
This in not completely unexpected: Damski \cite{Dam04} recently proposed
a `KZM approximation to LZF' in an insightful paper. However, even
upon closer inspection (work in progress~\cite{DDZ}), such agreement 
between LZF in the setting which is not a standard avoided level crossing
and the numerics might be still somewhat surprising.

When we compare the KZM and LZF predictions for defect density, we find:
\begin{equation}
\tilde \nu_{LZF} \simeq {1 \over {\tilde N}} = {1 \over {2 \pi}}  \sqrt{{2 |\ln(1-f)|} \over{\pi}  } \times \hat \nu_{KZM} \ .
\label{eq:ntildehat}
\end{equation}
The two estimates of defect density exhibit the same scaling with the quench rate
and with the parameters of $H$, Eq.~(\ref{eq:Hamiltonian}). However, LZF
predicts  fewer defects than the ``raw KZM estimate'' ($\tilde
\nu_{LZF} \simeq 0.14 \times \hat \nu_{KZM}$ when $f$ is set --
somewhat arbitrarily -- to 0.5). This is no surprise; numerical simulations, experiments and
analytic solutions to specific models, have shown that Eqs.~(\ref{eq:xihat})
and (\ref{eq:Nhat}) provide correct scalings, but tend to
overestimate densities (see e.g. \cite{LZ96a, Man03}). Fig.~1
confirms that this is also true also for the quantum Ising model.

We note that while $\hat \nu_{KZM}$ and $\tilde \nu_{LZF}$ are closely
related, they answer somewhat different questions. In particular,
$\hat \nu_{KZM}$ does not depend on $f$. However, when less than
one defect is expected in a chain, the number of defects
is $\simeq 1-f$ and can be computed using LZF. Fig.~3 shows that
LZF and KZM complement each other in this case, and jointly cover a
wide range of quench rates.

\begin{figure}[tp]
\centering
\includegraphics{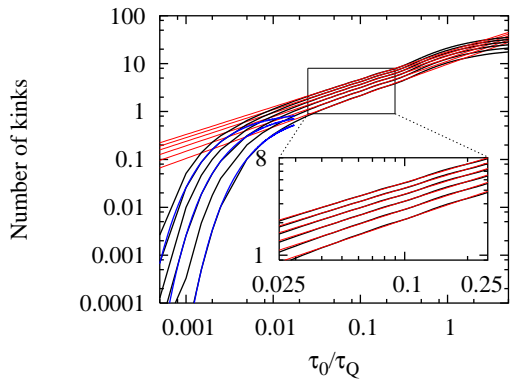}
\caption{%
  Number of kinks in chains ($N$=50,60,70,80,90,100, spins, bottom to
top) after a quench versus the quench rate
$\tau_0/\tau_Q = \hbar v/4W^2$. Both the scaling $\hat \nu \sim 1
/ \sqrt \tau_Q$ predicted by KZM (red lines), Eq.~(\ref{eq:Nhat}),
and the LZF estimate $1-f$ (blue lines)
are valid where expected. The red lines are linear fits in the range
(0.025,0.25) yielding slopes between 0.66 and 0.58. The blue lines
are the fit results from Fig.~\ref{lzf}(c).  Numerical data  (obtained using the method
developed in \cite{Dor03}) include these used in Fig. 1. However, we now go 
beyond the expected range
of validity of KZM: For quenches slow enough to create
`less than a kink' LZF provides
reliable predictions, while very fast quenches are 
``all impulse'', and -- according to KZM -- the number of kinks should saturate,
as is indeed seen.}
\label{lzfkzm}
\end{figure}

We found that a quantum analogue of KZM, based
on critical scalings, predicts the results of numerical
simulations. As expected, KZM scaling holds when $\hat \epsilon <
1$ -- i.e., when the quench starts and ends in the adiabatic regime, but
becomes impulse - like near the critical point. When the quench is
so slow that it never acquires impulse - like character, LZF is 
accurate.
We conclude that the two approaches work well in complementary
regimes of quench rates, and predict the same scaling for the size
of broken symmetry domains with quench time.

Exchanges of ideas with Robin Blume-Kohout and Bogdan Damski on
both the substance and the presentation are acknowledged and appreciated,
as is partial support by a Marie Curie Intra-European
Fellowship within the 6th European Community Framework Programme,
by a grant from NSA, and by the Austrian Science Foundation and EU Projects.

\end{document}